\documentclass[useAMS,usenatbib,epsf]{mn2e}
\begin{document}
\input{epsf}
\title[Integral field spectroscopy of IRAS 18276-1431]{Integral field
  spectroscopy of H$_2$ and CO emission in IRAS~18276-1431: evidence
  for ongoing post-AGB mass~loss}

\author[T.M. Gledhill et al.]
        {T.M.~Gledhill$^{1}$\thanks{email: {\tt t.gledhill@herts.ac.uk}},
         K.P.~Forde$^{1}$, K.T.E.~Lowe$^{1}$, M.D.~Smith$^{2}$\\
         $^{1}$Science and Technology 
         Research Institute, University of Hertfordshire, 
         College Lane, Hatfield AL10 9AB, UK \\
         $^{2}$Centre for Astrophysics \& Planetary Science, School of Physical 
         Sciences, University of Kent, Canterbury CT2 7NR, UK
         }
\maketitle

\begin{abstract}
We present {\it K}-band integral field spectroscopy of the bipolar
post-AGB object IRAS 18276-1431 (OH 17.7-2.0) using {\sc SINFONI} on
the {\sc VLT}.  This allows us to image both the continuum and
molecular features in this object from $1.95-2.45\umu$m with a spatial
resolution down to 70~mas and a spectral resolution of $\sim 5000$.
We detect a range of H$_2$ ro-vibrational emission lines which are
consistent with shock excitation in regions of dense ($\sim 10^7$~cm$^{-3}$) gas
with shock velocities in the range $25-30$~km~s$^{-1}$. The
distribution of H$_2$ emission in the bipolar lobes suggests that a
fast wind is impinging on material in the cavity walls and tips. H$_2$
emission is also seen along a line of sight close to the obscured star
as well as in the equatorial region to either side of the stellar
position which has the appearance of a ring with radius
0.3~arcsec. This latter feature may be radially cospatial with the boundary
between the AGB and post-AGB winds. The first overtone $^{12}$CO
bandheads are observed longward of 2.29~$\umu$m with the ${\rm v}=2-0$
bandhead prominently in emission. The CO emission has the same spatial
distribution as the {\it K}-band continuum and therefore originates
from an unresolved central source close to the star. We interpret this
as evidence for ongoing mass loss in this object. This conclusion is
further supported by a rising {\it K}-band continuum indicating the
presence of warm dust close to the star, possibly down to the
condensation radius. The red-shifted scattered peak of the CO bandhead
is used to estimate a dust velocity along the bipolar axis of
95~km~s$^{-1}$ for the collimated wind. This places a lower limit of
$\sim 125$~yr on the age of the bipolar cavities, meaning that the
collimated fast wind turned on very soon after the cessation of AGB
mass loss.
\end{abstract}


\section{Introduction}

Imaging surveys of young post-asymptotic giant branch (post-AGB)
stars, not yet hot enough to ionize their envelopes, show that many
are already associated with complex nebulosities. They exhibit a range
of structural symmetries, such as bipolar, multipolar and
point-symmetric (e.g. Si\'{o}dmiak et al. 2008, Sahai et al. 2007,
and references therein). These objects then evolve into ionized
planetary nebulae (PNe) with similar morphologies which strongly
suggests that (i) the shaping process operates and is largely
completed during this earlier pre-PN phase and (ii) the onset of
shaping may be linked with the appearance of a collimated fast wind
early in the post-AGB phase, which then carves axisymmetric channels
into the remnant AGB envelope (e.g. Sahai \& Trauger 1998; Lee, Hsu \&
Sahai 2009).

The appearance of fast collimated winds at the end of the AGB is 
linked with the
emergence of collisionally excited molecular transitions (e.g. Davis
et al. 2005).  In particular, the {\it K}-band ro-vibrational
transitions of H$_2$ will be excited, and these provide an important
tracer of shocked gas in pre-PNe. Imaging and spectroscopic surveys
(e.g.  Hrivnak, Kwok \& Geballe 1994; Garc\'{i}a-Hern\'{a}ndez et
al. 2002; Davis et al. 2003, 2005; Kelly \& Hrivnak 2005) show that,
for spectral types later than B, the stellar UV flux is still not
usually sufficient for radiative excitation to dominate and so the
H$_2$ emission is produced in shocks. There appears to be a
strong correlation between shock-excited H$_2$ and bipolarity,
something that has also been noted for PNe (Kastner et al. 1996).

IRAS 18276-1431 (OH 17.7-2.0) is a fairly well studied post-AGB object
and a review of its properties is given by Bains et al. (2003). These
authors present interferometry and polarimetry of the OH maser lines,
showing evidence for a molecular torus expanding slowly at
13~km~s$^{-1}$, threaded by an ordered magnetic field. In the near-IR
the object has a bipolar appearance and imaging polarimetry in the
{\it K}-band shows that the lobes are seen in scattered light with the
source remaining hidden (Gledhill 2005). High angular resolution
imaging in the $K_{p}$-, $L_{p}$- and $M_{s}$-bands reveals the
structural details of the lobes and halo, including searchlight beams
and concentric multiple arcs (S\'{a}nchez Contreras et al. 2007;
hereafter SC07).  These authors also present CO interferometry
observations and a spherically symmetric radiative transfer model of
the spectral energy distribution (SED).

The near-IR continuum observations provide information on the
distribution of dust and optical depth throughout the object; the
bipolar structures nested within a more spherical halo suggest the
recent development of axisymmetry linked to the onset of a
collimated post-AGB wind.  In this paper we concentrate on the
molecular emission, its excitation and the interaction between any
collimated outflow and the AGB envelope in IRAS 18276-1431 (hereafter
IRAS 18276). We present integral field spectroscopy
using the SINFONI instrument on VLT, which allows us to map the
distribution of the various {\it K}-band molecular emission features
over the object.

\section{Observations and data reduction}

Observations of IRAS 18276 were made on 2005 June 30 with the SINFONI
integral field spectrometer on the 8.2-m UT4 telescope at the VLT
observatory in Chile (Eisenhauer et al. 2003, Bonnet et al. 2004). The
{\it K}-band grating was used covering a wavelength range of $1.95$ to
$2.45~\umu$m with spectral resolution of $\sim 5000$. This equates to
a spectral pixel (channel) width of $2.45\times10^{-4}~\umu$m.  The
expected spectral resolution is 2 channels or $4.9\times10^{-4}~\umu$m
($\approx 66$~km~s$^{-1}$ at $2.2~\umu$m).  Both Medium Resolution Mode
(MRM) and High Resolution Mode (HRM) observations were made.

In MRM mode, the field of view is 3$\times$3 arcsec with
$50\times100$~mas spatial pixels. A total of 6 integrations were made,
at two overlapping pointings, so that the total integration time on
IRAS 18276 was $6\times200$~sec = 20~min. Offset sky exposures were
obtained along with the standard arc, flux and telluric
calibrations. Flux and telluric calibration was performed using
HIP~85920. The airmass range of these observations was $1.023-1.060$
with an ambient seeing of $\approx 0.8$ arcsec. AO-correction in MRM
resulted in an average PSF of 150~mas FWHM (measured using the
field star FS1 - see Fig.~1a).

HRM provides a field of view of $0.8\times 0.8$~arcsec with
$12.5\times25$~mas pixels. The HRM observations were obtained at high
airmass at the end of the night as the object was setting. Telluric
correction of these observations proved difficult and in the end we
rejected all but 3 data sets, where the airmass was less than 1.9. A
single pointing was used in these observations so that the total
integration time on-source is $3\times300$~sec = 14~min.  HIP~18302
was used as the flux and telluric standard. Ambient seeing was again
$\approx 0.8$~arcsec and AO correction resulted in a PSF width of $\approx
70$~mas FWHM estimated from the standard star.

The MRM and HRM data were reduced in a common manner using the ESO
pipeline for SINFONI with further processing and data visualization
using the STARLINK software collection.

\section{Results}
\subsection{White light images}
\begin{figure}
\epsfxsize=8.5cm \epsfbox[10 -30 262 633]{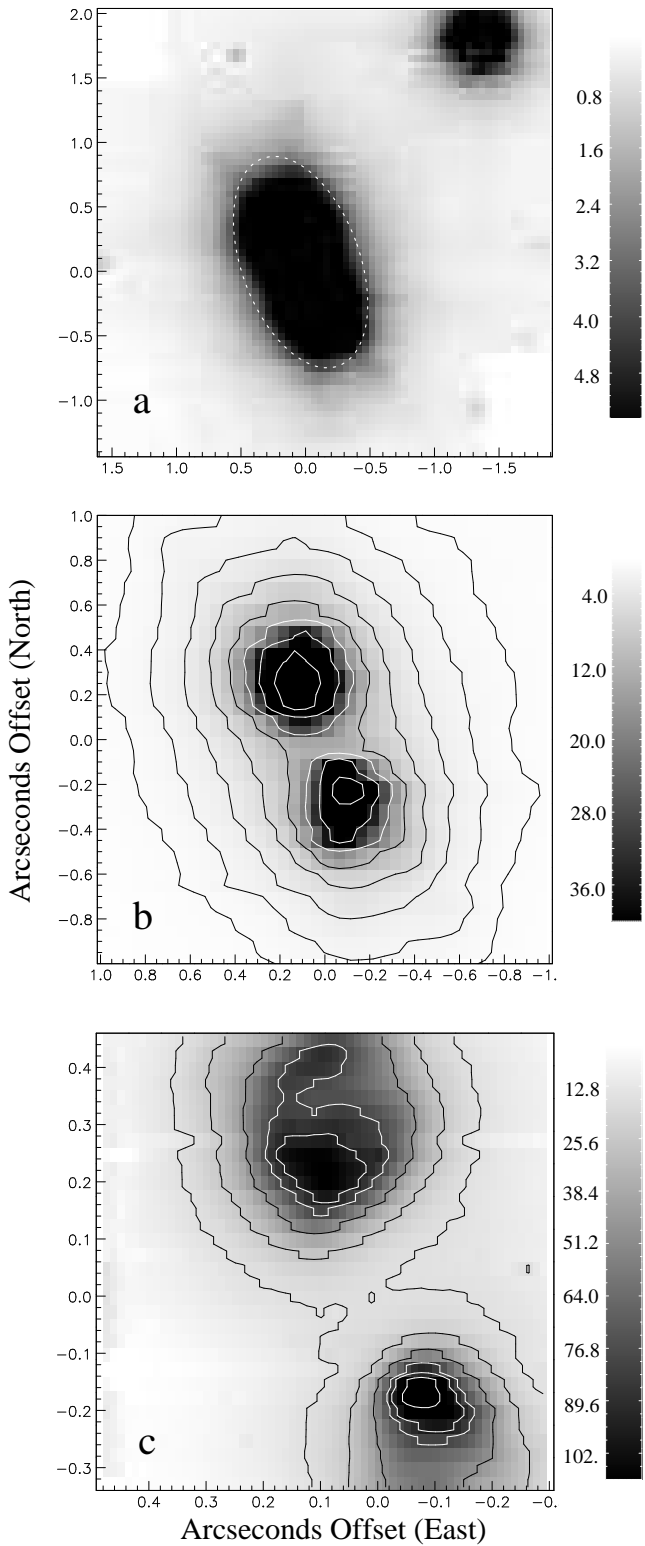}
\vspace{-2cm}
\caption{{\it K}-band `white light' images of IRAS 18276. MRM 
images show the
  faint {\it(a)} and bright {\it(b)} emission. Contours in {\it (b)}
  are at 0.8, 1.5, 2.9, 5.5, 10.6, 20.1, 38.3, 72.9. The dotted
  ellipse shows the region over which the {\it K}-band spectrum (Fig.~2) is
  extracted. The star to the top right in {\it (a)} is the field star FS1
(see text). In {\it (c)} we show the HRM image. 
  Contour levels in this image are at 15.4, 25.6,
  47.4, 79.4, 94.7, 126.7. All greyscale and contours levels are in
  units of 10$^{-15}$~W~m$^{-2}$~arcsec$^{-2}$.}
\end{figure}

The SINFONI spectral datacube can be collapsed along the
wavelength axis to produce a `white light' image. In Fig.~1 we show
the MRM and HRM data summed over the range 1.95--2.45~$\umu$m, to form
white light images of the object, dominated by the scattered
continuum emission. The upper panel (Fig.~1a) displays the entire
SINFONI field, comprising two telescope pointings, and shows the
faint emission levels. The star to the upper right was identified as a
field star, not physically associated with IRAS 18276, in {\it J}- and
{\it K}-band polarimetric observations (Gledhill 2005) and is the star
labelled `FS1' by SC07. The four searchlight beams identified by
SC07 are visible, extending through the faint halo.

In the central panel (Fig.~1b), the brighter continuum emission is
shown, revealing two lobes either side of the central position, forming
a bipolar nebula. Polarimetric observations show high degrees of linear
polarization in the lobes ($\approx 50$ per cent in the {\it K}-band) 
confirming  that the nebulosity
is seen by scattered light (Gledhill 2005). The star
is hidden by a circumstellar torus of dust and gas completely
obscuring the direct view at these wavelengths. SC07
estimate a lower limit to the optical depth of the circumstellar
obscuration at 2.12~$\umu$m of $\tau_{2.12} > 12.3$. We define the
origin of our coordinate system in these images as the centre of the
obscuring lane along a line joining the brightness peaks of the two lobes.

The HRM image, shown in Fig.~1c, provides a field of view of
$0.8\times0.8$ arcsec and close to diffraction-limited imaging on the
8.2-m VLT. The origin of the coordinate system is located as before,
at the centre of the obscuring lane midway between the two lobes. The
structure in the HRM image closely resembles that seen in the sharpened
{\it K$_p$} Keck telescope image of SC07, with the limb-brightened tip or
`cap' clearly visible in the northern lobe. The axis defined by the
two continuum peaks lies at a position angle (PA) of 24\degr.

The {\it K}-band flux density is $88\pm3$~mJy (the error mainly coming from
the telluric correction), corresponding to $K=9.67\pm0.04$
mag. This agrees reasonably with the flux density of 96~mJy quoted by
SC07 for observations just 1 month after our own. There is also
agreement within errors with the $K=9.5\pm0.3$ measurement of July
2001 (Gledhill 2005).

\subsection{Integrated spectrum}
The spectrum of IRAS 18276, from 1.95--2.45~$\umu$m, after telluric
correction and flux calibration, is shown in Fig.~2, and is integrated
over the elliptical aperture shown in Fig.~1a. The continuum rises
smoothly throughout the {\it K}-band and originates mainly from the
bipolar lobes, where light from the hidden central source is scattered
into our line of sight by dust grains. Superimposed on the continuum
there are numerous emission features corresponding to the S- and
Q-branch ro-vibrational transitions of H$_2$, which are labelled. We
label the first four $^{12}$CO overtone bandheads, which are
prominent in emission longward of 2.29~$\umu$m. We also detect the
Na~{\sevensize I} doublet lines in emission around 2.2~$\umu$m. 
We do not detect
Br$\gamma$. Other emission features in the spectrum result from
incomplete correction for metal absorption lines in the telluric
standard.

\subsection{Continuum subtraction and line flux estimation}
In order to get a good fit to the continuum in the vicinity of each
emission line, the {\it K}-band spectral cube was sectioned, in the
wavelength direction, into smaller `line cubes', each containing a
single emission line and surrounding continuum. A linear fit was then
made to the continuum on either side of the line, extrapolating under
the line. The {\sevensize MFITTREND} application, part of the
STARLINK software collection, was used to perform this
fit along the wavelength direction for each spatial pixel in the line
cube, producing a continuum cube. The continuum cube was then
subtracted from the line cube to leave a continuum-subtracted spectral
cube containing the emission line.

To estimate the flux in a line, aperture photometry using an
elliptical aperture was performed for each spectral image (channel)
in the cube containing flux above the background noise. The fluxes
estimated for each H$_2$ line and for the ${\rm v}=2-0$~CO bandhead emission
are given in Table~1 along with the rest
and observed wavelengths of the lines. The H$_2$ line peaks are
consistently offset to the red of the rest wavelength, with an average
offset of $4\pm1 \times 10^{-4}$~$\umu$m (a wavelength channel is
$2.45\times 10^{-4}$~$\umu$m) whereas the CO bandhead peak is offset to
the red by $1.3 \times 10^{-3}$~$\umu$m. The velocity structure of the
emission is discussed further in Section~\ref{velocity}.

\begin{figure*}
\epsfxsize=18.5cm \epsfbox{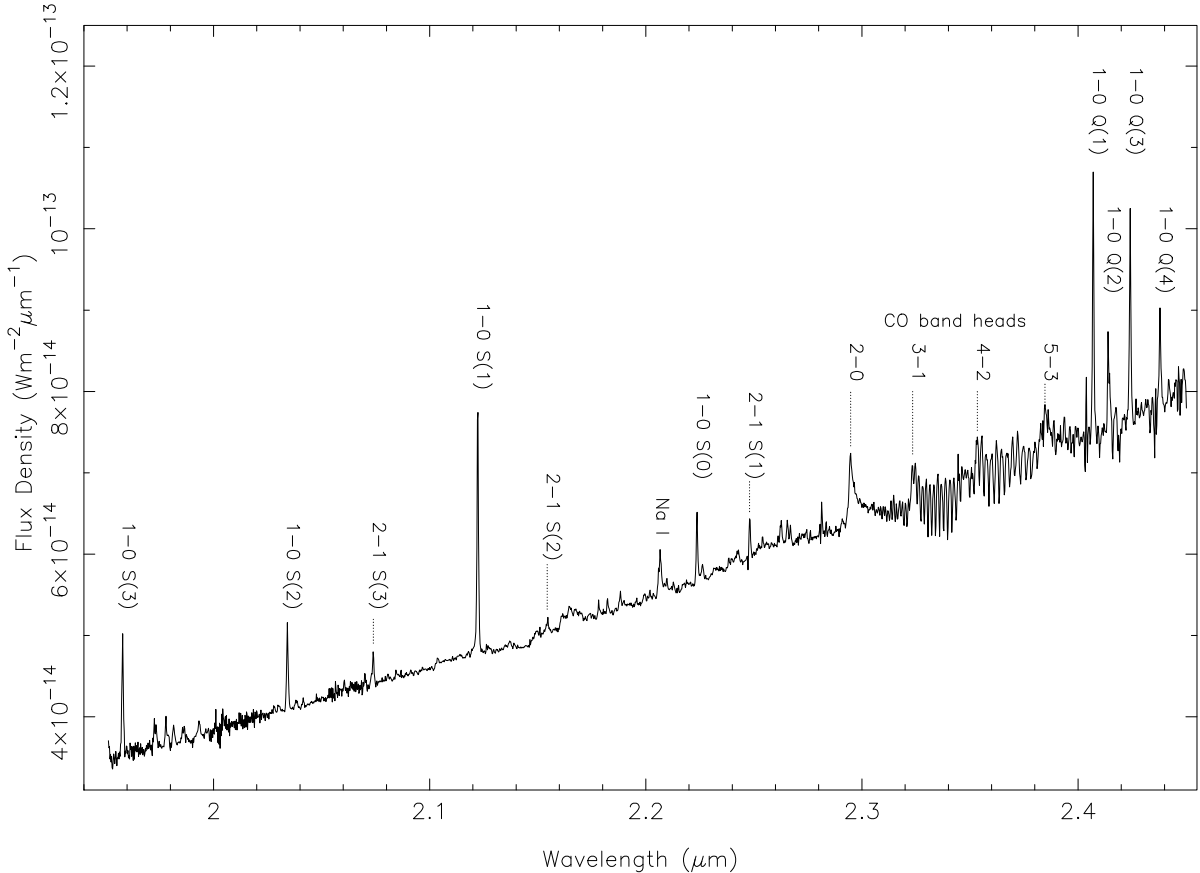}
\caption{The integrated spectrum of IRAS 18276 between 1.95 and 2.45~$\umu$m, 
showing the $1-0$ and $2-1$ H$_2$
ro-vibrational lines, CO first overtone bandheads and Na~{\sevensize I}
lines. Other
  emission features result from residual metal absorption lines in the
  telluric standard. }
\end{figure*}

\begin{table}
\caption{Continuum-subtracted line fluxes for the H$_2$ lines, along
  with their rest ($\lambda_{0}$) and measured ($\lambda_{p}$) peak wavelengths ($\pm
  0.0001$~$\umu$m). We also give the flux in the ${\rm v}=2-0$ CO bandhead
  emission. The fluxes have not been corrected for any line-of-sight
  extinction (see Section~3.4).}
\label{line-fluxes}
\begin{tabular}{lccc}
\hline 
Line    &$\lambda_{0}$ ($\umu$m)&$\lambda_{p}$ ($\umu$m)&F $\times 10^{-18}$ (W~m$^{-2}$) \\
\hline
1-0 S(3) & 1.9576 & 1.9580 & 10.65 $\pm 1.59$  \\
1-0 S(2) & 2.0338 & 2.0341 & 7.22  $\pm 1.19$  \\
2-1 S(3) & 2.0735 & 2.0739 & 2.36  $\pm 1.30$  \\
1-0 S(1) & 2.1218 & 2.1221 & 23.51 $\pm 1.57$  \\
2-1 S(2) & 2.1542 & 2.1543 & 0.89  $\pm 1.16$  \\
1-0 S(0) & 2.2235 & 2.2238 & 5.30  $\pm 1.18$  \\
2-1 S(1) & 2.2477 & 2.2480 & 2.51  $\pm 1.35$  \\
1-0 Q(1) & 2.4066 & 2.4070 & 20.46 $\pm 1.87$  \\
1-0 Q(2) & 2.4134 & 2.4139 & 7.26  $\pm 1.89$  \\
1-0 Q(3) & 2.4237 & 2.4242 & 17.68 $\pm 1.98$  \\
1-0 Q(4) & 2.4375 & 2.4379 & 6.65  $\pm 1.92$  \\
2-0 CO bh& 2.2935 & 2.2948 & 46.6  $\pm 3.7$   \\
\hline
\end{tabular}
\end{table}

\subsection{Extinction correction}
In principle it is possible to estimate the amount of extinction
encountered by the H$_2$ emission along its path by comparing certain
$1-0$ Q- and S-branch line strengths. In particular, the Q(4)/S(2),
Q(3)/S(1) and Q(2)/S(0) line pairs share the same upper rotational
level and the measured line strength ratios therefore depend only on
known constants plus the differential extinction suffered by the two
lines.  Although this technique has been widely used to estimate
extinction (e.g. Davis et al. 2003; Chrysostomou et al. 1993), in
practice it is problematic, largely due to the Q-branch lines lying
beyond 2.4~$\umu$m in a region of poor and variable atmospheric
transmission.

All three lines pairs are present in our data, allowing three
extinction estimates. We give the resulting ratios and extinction
values at the wavelength of the $1-0$~S(1) line in Table~2. As can be
seen there is considerable spread within the errors. The most reliable
estimate should be the Q(3)/S(1) ratio, as this combines the two
strongest lines. The Q(2)/S(0) ratio is suspect as there is some
evidence that the Q(2) line is contaminated with the 6-4 $^{12}$CO
bandhead at 2.4142~$\umu$m. The Q(4)/S(2) ratio combines two
relatively weak lines, with the Q(4) line being at the long wavelength
limit of our spectral range. SC07 derive an interstellar extinction of
$A_{\rm v}=1.6\pm0.5$ mag. to IRAS 18276, which they equate to 0.3
mag. at 2.2~$\umu$m. This combined with the Q(3)/S(1) ratio suggests
that the {\it K}-band extinction to the H$_2$ emitting region lies 
somewhere in the range 0.3--1
mag.  However, in view of the uncertainties, we have not
extinction-corrected the line fluxes in Table~1.

\begin{table}
\caption{Line pairs that can be used to estimate the {\it K}-band
  extinction. $R$ is the expected (unattenuated) line ratio, and
  $R^{\prime}$ the measured ratio and error. $A_{2.122~\umu{\rm m}}$ is
  the corresponding extinction at 2.122~$\umu$m and the final column
  gives the range of extinction allowed by the error on
  $R^{\prime}$. }
\label{extinction}
\begin{tabular}{lcccc}
\hline 
Lines    & $R$ & $R^{\prime}$ & $A_{2.122~\umu{\rm m}}$ (mag) & Range (mag) \\
\hline
Q(3)/S(1) & 0.70   & 0.75$\pm0.10$ &  0.38  & 0 $\rightarrow$ 1.0 \\
Q(2)/S(0) & 1.10   & 1.37$\pm0.51$ &  1.37  & 0 $\rightarrow$ 4.7  \\
Q(4)/S(2) & 0.56   & 0.92$\pm0.31$ &  1.84  & 0.3 $\rightarrow$ 2.9 \\
\hline
\end{tabular}
\end{table}

\section{Evidence for H$_2$ shock excitation}

The H$_2$ ro-vibrational lines in post-AGB objects can be excited
collisionally by shocks (for example in outflows and winds) or
radiatively by stellar UV photons, or by a combination of both
mechanisms. Collisional excitation will tend to populate the lower
vibrational levels first, whereas UV-excitation leads to population of
the higher ${\rm v}\ge3$ levels. A comparison of emission lines from lower
and higher states can be used to distinguish between the excitation
mechanisms. 

\subsection{Line ratios}
The $1-0$~S(1)/$2-1$~S(1) ratio is a commonly used diagnostic of H$_2$
excitation, the typical shock value being $\approx 10$ although a wide
range of values is possible, from $\approx 4$ upwards, depending on the
pre-shock gas density and shock velocity (e.g. Shull \& Hollenbach
1978; Smith 1995). In the case of pure radiative excitation by UV
photons the ratio is $\approx 2$ (Black \& Dalgarno 1976). For IRAS 18276
we calculate $1-0$~S(1)/$2-1$~S(1)~=~$9.4\pm3.2$, averaged over the
object (Table 1), consistent with shock excitation. Inclusion of
differential extinction increases this ratio slightly, to 10.5 and
11.8 for a {\it K}-band (2.2~$\umu$m) extinction of 1 and 2 mag.,
respectively\footnote{Here we have assumed that extinction in the infrared
varies as $A_{\lambda} \propto \lambda^{-\alpha}$, with
$\alpha=2.14$. This is steeper than the normally assumed Galactic
extinction law ($\alpha = 1.6-1.8$) and has recently been determined
from UKIDSS Galactic Plane Survey data (Stead \& Hoare, 2009).}.

In regions of high-density gas ($>10^{5}$~cm$^{-3}$), collisional
  de-excitation of the ${\rm v}=2$ vibrational level can lead to a
  $1-0$~S(1)/$2-1$~S(1) line ratio which mimics that of shock-excited
  gas even in cases of pure radiative excitation (e.g. Hollenbach \&
  Natta 1995). There are clearly regions of high density in IRAS 18276
  and SC07 estimate a gas density of $2-4 \times 10^{7}$~cm$^{-3}$ in
  the lobe caps from the infrared colours. A useful excitation
  diagnostic for high-density regions is provided by the
  $1-0$~S(1)/$3-2$~S(3) line ratio; for high densities
  ($10^{6-7}$~cm$^{-3}$) this ratio maintains a value $\approx 8$ for
  UV-excited H$_2$ emission, for UV fluxes of $\sim 10^3$ times the
  ambient interstellar radiation. Only in the case of very UV-intense
  photodissociation regions does it approach shock values (Burton,
  Hollenbach \& Tielens 1990).  We note a weak feature in our spectrum
  at the expected location of the $3-2$~S(3) line and estimate an
  upper flux limit of $3 \times 10^{-19}$~W~m$^{-2}$, placing a lower
  limit on the $1-0$~S(1)/$3-2$~S(3) line ratio of $\approx 80$.  This
  is consistent with any $3-2$~S(3) flux being produced in shocks
  (e.g.  Smith 1995), suggesting little if any UV excitation. This
  accords with expectation if the star has effective temperature
  $T_{\rm eff}=7000$~K, as modelled by SC07. Studies of H$_2$ emission
  in post-AGB objects suggest that radiative excitation is not
  prominent in stars later than B spectral type (Kelly \& Hrivnak
  2005; Garc\'{i}a-Hern\'{a}ndez et al. 2002).

\subsection{The ortho-para ratio}

The ortho-para ratio (OPR) can also be used to distinguish between
shock and radiative excitation. At the point of formation on dust
grains, the H$_2$ OPR is assumed to be 3, reflecting the ratio of spin
degeneracies for ortho (odd-J) and para (even-J) states. Subsequently,
various processes can lead to a lower OPR, as detailed by Martini,
Sellgren \& Hora (1997).  In particular, UV excitation lowers the OPR,
with values as low 1.8 and 1.7 measured in photodissociation regions
such as M17 (Chrysostomou et al. 1993) and the planetary nebula Hb12
(Ramsay et al.  1993). In contrast, where collisional excitation
dominates, the OPR is expected to remain close to 3. 

We follow the prescription of Smith, Davis \& Lioure (1997) for 
calculating the ${\rm v}=1$ OPR, $\phi$, from the $1-0$~S(0), S(1) and S(2) lines:

\begin{equation}
\phi = 0.809\eta \left[\frac{F_{1}}{F_{2}}\right]^{0.431}
\left[\frac{F_{1}}{F_{0}}\right]^{0.569}
\end{equation}

\noindent where $F_{0}$, $F_{1}$, $F_{2}$ are the line fluxes and $\eta$ is
determined by the relative extinction of the lines. In the case of the
three lines used here $\eta=1.004$ and we find, using the integrated
line fluxes from Table~1, $\phi=3.02\pm0.18$.

The continuum-subtracted spectral cube can be collapsed in wavelength
over each line to form line images, which can then be used with
Equation~1 to form the OPR image, as shown in Fig.~3. There is little
evidence for variation in the OPR over the object and, within the
errors, the H$_2$ emission is consistent with an OPR of 3. This again
supports the contention that the H$_2$ emission in IRAS 18276 is
shock-excited and that UV-pumped fluorescence is not present.

\begin{figure}
\label{fig_opr}
\epsfxsize=8.5cm \epsfbox{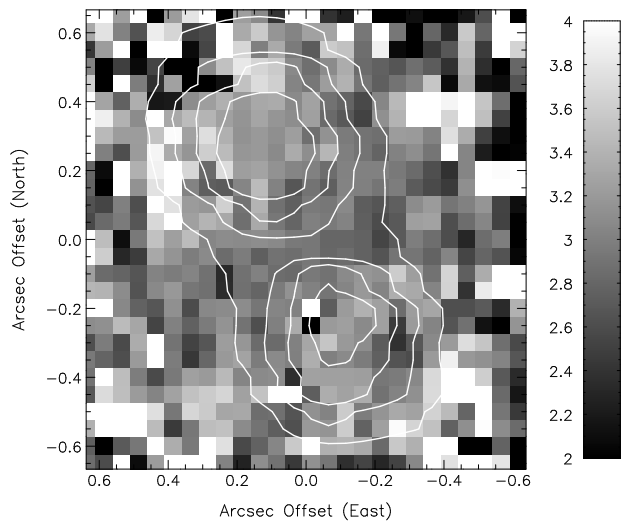}
\caption{The ortho-para ratio image, calculated from the $1-0$~S(0,1,2)
  lines, superimposed with contours of {\it K}-band surface brightness (see
  Fig.~1) to indicate the location of the nebulosity.}
\end{figure}

\subsection{Rotational and vibrational temperatures}

In regions where shocks dominate, the level populations are expected
to be thermalised, resulting in similar rotational and vibrational
temperatures: $T_{\rm rot}\sim T_{\rm vib}$ (see Burton 1992 for a
review). In UV-dominated regions, the higher vibrational levels (${\rm v}\ge
2$) are over-populated so that $T_{\rm vib} \gg T_{\rm rot}$.  A
comparison of the rotational and vibrational temperatures can
therefore be used as an indication of the excitation mechanism
(e.g. Tanaka et al. 1989).

A convenient graphical method to determine the gas temperature,
$T_{\rm g}$, is to plot normalised column densities against upper
level temperatures (e.g. Hasegawa et al. 1987; Martini, Sellgren \&
Hora 1997; Rudy et al. 2002):

\begin{equation}
\ln \left[\frac{N_{i}g_{(1,3)}}{N_{(1,3)}g_{i}}\right]T_{\rm g} = T_{i}-T_{(1,3)}
\end{equation}

\noindent where $N_i$ is the column density of H$_2$ in the upper level, $i$, of
a particular transition, and $N_{(1,3)}$ is the upper level column
density for the $1-0$~S(1) line (${\rm v}=1, {\rm J}=3$). The respective statistical
weights are $g_i$ and $g_{(1,3)}$ and we use the ortho-para ratio of
3.0 calculated in the previous section. In such a plot, $T_{\rm g}$ is
given by the inverse of the slope of a line connecting the data
points. We derive two estimates of the rotational temperature, $T_{\rm
  rot}$, for the $1-0$ and $2-1$ transitions, by fitting a line through
transitions within the same vibrational level. The vibrational
temperatures, $T_{\rm vib}$, are obtained by comparing ${\rm v}=2-1$ and
${\rm v}=1-0$ transitions with the same rotational quantum number, J. The
column densities, $N_i$, are obtained from the line fluxes, $F_i$, in
Table~1, where:

\begin{equation}
F_i=\frac{hcN_iE_i}{\lambda_i}10^{-0.4A_{i}}
\end{equation}

\noindent with $\lambda_i$ the wavelength, $E_i$ the quadrupole transition
probability (Turner, Kirby-Docken \& Dalgarno 1977) and $A_{i}$ the 
extinction, for the line. 

\begin{figure}
\epsfxsize=11cm \epsfbox[100 82 416 500]{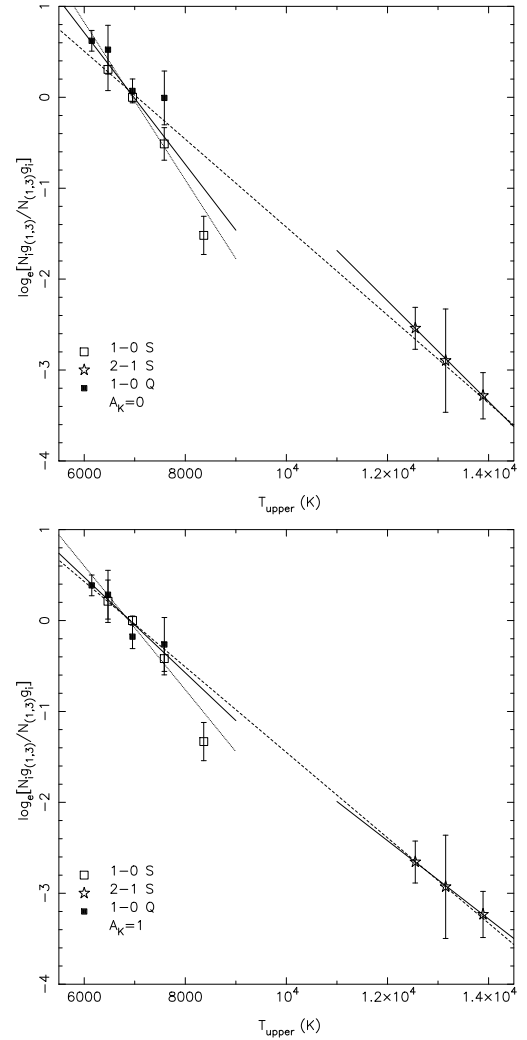}
\caption{The log of the normalised column densities versus upper level
  temperature for the H$_2$ emission lines in Table~1. The upper plot
  is not corrected for extinction whereas the lower plot has an extinction 
  correction of $A_{K}=1$~mag. applied.  Separate fits to the ${\rm v}=1-0$ 
  points [excluding $1-0$~S(3)]  and ${\rm v}=2-1$ points are
  shown as solid black lines, and give estimates of $T_{\rm rot}$. A
  fit through the combined $1-0$ and $2-1$ data points [excluding $1-0$~S(3)] 
  is shown by a dashed line. A fit through all the $1-0$ points is shown
 as a light black line.  }
\end{figure}

\begin{table}
\caption{Rotational and vibrational temperatures calculated from the
  normalized column densities plotted in Fig.~4. Excluding the
  $1-0$~S(3) point the fits are consistent with $T_{\rm
    rot} \approx T_{\rm vib}$.  A fit through all the data points gives
  the temperature, $T_{\rm g}$, that the gas would have if
  characterized by a single excitation temperature.}
\label{temps}
\begin{tabular}{lcc}
\hline 
Estimate                 &\multicolumn{2}{c}{Temperature (K)}         \\
                         &       $A_{K}=0$        &    $A_{K}=1$       \\
$T_{\rm rot}$ ${\rm v}=1-0$     & $1379\pm226$           & $1901\pm431$       \\
$T_{\rm rot}$ ${\rm v}=2-1$     & $1805\pm 834$          & $2320\pm1378$      \\
$T_{\rm vib}$ (J=3)       & $2201\pm202$           & $2106\pm185$       \\
$T_{\rm vib}$ (J=4)       & $2334\pm570$           & $2216\pm514$       \\
$T_{\rm g}$                  & $2053\pm123$           & $2128\pm132$       \\\\
$T_{\rm rot}$ ${\rm v}=1-0^{\dagger}$     & $1155\pm122$  & $1468\pm197$       \\
$T_{\rm vib}$ (J=5)$^{\dagger}$     & $3132\pm522$  & $2904\pm449$       \\
\hline
\multicolumn{2}{l}{$\dagger$ fits including the $1-0$~S(3) point} \\
\end{tabular}
\end{table}

We plot our data in Fig.~4, both with no extinction correction (top)
and with a correction for a {\it K}-band ($\lambda = 2.2~\umu$m)
extinction of 1 mag. (lower), which is the maximum value consistent
with the Q3/S1 ratio (Sec.~3.4). The extinctions at each transition
wavelength, $A_{i}$, are extrapolated assuming a power-law
relationship (see footnote 1). We note that the $1-0$~S(3) point, at
an upper level temperature of 8365~K, appears rather low in these
plots. Although the line appears strong in Fig.~2, its flux relative
to the $1-0$~S(1) line is in fact much weaker than expected for
shock-excited gas: from our data $1-0$~S(1)/$1-0$~S(3) = 2.21 whereas a
C-shock model from Smith (1995) gives a ratio of 1.13. The $1-0$~S(3)
line lies outside the {\it K}-band, at 1.96~$\umu$m, in a region of
poor atmospheric transmission, and may therefore be subject to a poor
telluric correction. In view of these doubts, we have performed fits
to the data in Fig.~4 both including and excluding the
$1-0$~S(3) line.

The solid black lines show weighted least-squares fits to the $1-0$
and $2-1$ data points, excluding the S(3) point, providing two
estimates of $T_{\rm rot}$. Correcting for 1 mag.~of extinction (lower
panel) improves the fit, reducing the scatter on the $1-0$ data
points. Increasing the extinction correction much beyond $A_{\rm K}=1$
starts to increase the scatter again. These fits indicate a rotational
temperatue of $\approx 2000$~K within errors for the extinction
corrected data (Table~3).
 
The three estimates of $T_{\rm
  vib}$ are formed by comparing the three ${\rm v}=2-1$ rotational levels
with their ${\rm v}=1-0$ equivalents. The resulting temperatures and errors
are given in Table~3.  We can see that the vibrational temperature
based on the $2-1$~S(3) and $1-0$~S(3) lines (J=5) appears anomalously
high, which again casts doubt on our measured $1-0$~S(3) flux. In fact, a
reasonable straight line fit can be made to all the data points if the 
$1-0$~S(3) point is excluded, and this is shown as a
dashed line in Fig.~4. This would imply that $T_{\rm rot} \approx T_{\rm
vib}$, which is typical of shock-excited gas (Tanaka et al. 1989).

\subsection{Shock models}
\label{shocks}

\begin{figure*}
\epsfxsize=17cm \epsfbox{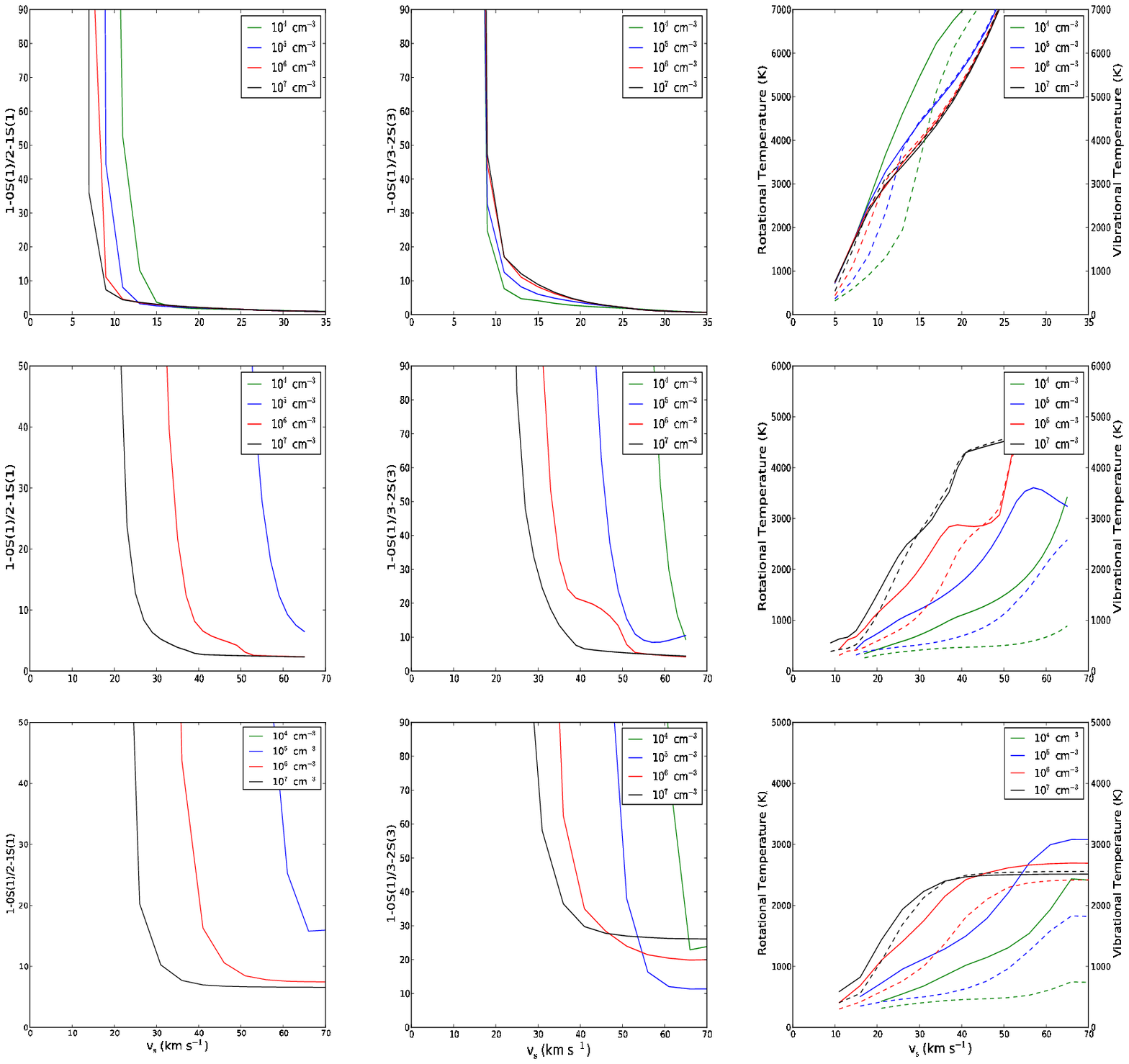}
\caption{Model plots showing J-type planar (upper), C-type planar
(middle) and C-type bow (lower) shocks. From left to right the panels
show the $1-0$~S(1)/$2-1$~S(1) line ratio, the $1-0$~S(1)/$3-2$~S(3) line
ratio and the excitation temperatures (rotational as solid lines and
vibrational as dotted lines) as a function of shock velocity, and for
a range of gas densities according to the legend in the inset boxes. See main
text for further details. A colour version of this Figure is available
online.}
\end{figure*}

The velocity of the shock required to collisionally populate the ${\rm
  v}=2$ vibrational level depends on the density of the gas into which
the shock propagates. For a planar C-shock with a transverse magnetic
field (maximum cushioning) and a pre-shock gas density of
$10^{7}$~cm$^{-3}$, a shock velocity of $\approx 20$~km~s$^{-1}$ can
produce the observed $1-0$~S(1)/$2-1$~S(1) line ratio of $\approx 10$ (Le
Bourlot et al. 2002). SC07 estimate a gas density (based on dust
optical depth) of $2-4\times10^{7}$~cm$^{-3}$ in the limb-brightened
edges and cap of the northern lobe, so that a shock in this velocity
range could produce the observed H$_2$ emission in this region. These
shocks, propagating into the cavity walls, may also contribute to the
overall envelope expansion. Spatially unresolved $^{12}$CO ${\rm J}=1-0$
observations give an average envelope expansion velocity of
17~km~s$^{-1}$ (SC07). This is similar to the 13~km~s$^{-1}$ expansion
velocity determined from OH maser observations (Bains et al. 2003),
which trace the kinematics within a 1 arcsec radius of the star.

To further refine these arguments we calculate the expected line
ratios [$1-0$~S(1) relative to $2-1$~S(1) and $3-2$~S(3)] and
rotational and vibrational temperatures for various shock velocities
and gas densities.  We consider J- and C-type planar and curved (bow)
shocks, using the models described by Smith, Kanzadyan \& Davis
(2003), Smith (1994), Smith \& Brand (1990) and references therein. In
the case of bow shocks, a paraboloidal shock is used with the shape
characterised by a curvature exponent $s=2$ (see Smith et al. 2003).
We assume that all of the hydrogen is in molecular form. 

We find that planar J-shocks, Fig.~5 (upper), produce
too much flux in the ${\rm v}=2$ and ${\rm v}=3$ lines to remain
consistent with our observed line ratios, except for shock velocities
less than 8~km~s$^{-1}$.  This velocity limit seems unreasonably low given the
evidence for bulk motions of the molecular gas in excess of this
figure, from CO and OH measurements, and the likelihood of a faster
wind component (discussed in Section~\ref{COvelocity}). This is also
true of J-type bow shocks, although a velocity of up to 10~km~s$^{-1}$
is allowed in this case. Higher velocity planar J-shocks will also
raise the gas temperature to $\sim 3000$~K, which again is
inconsistent with our line ratios.

C-shocks occur when a magnetic field cushions the shock in gas with a
low pre-shock ionization fraction, typical of molecular clouds. If the
ionization fraction exceeds $\sim 10^{-5}$ then a J-shock occurs
(Smith 1994), however in the case of IRAS 18276 there is no evidence
for an ionized region.  The detection of Zeeman-split OH maser
components in the equatorial regions of IRAS 18276 argues for a
magnetic field of a few mG in these regions (Bains et al. 2003).
Taking $B = 4$~mG, then for a gas density of $10^{7}$~cm$^{-3}$, the
Alfv\'{e}n speed is $V_{a}\approx 2$~km~s$^{-1}$ [given by
  $V_{a}=1.85\times10^{3}B({\rm mG})n({\rm
    cm}^{-3})^{-0.5}$~km~s$^{-1}$; e.g.  Smith 1994]. The C-shock
models for $V_{a}=2$~km~s$^{-1}$ are shown in Fig.~5 (middle) and
(lower) for planar and bow shock respectively. A common characteristic
of the C-shocks is that they predict significantly different
rotational and vibrational temperatures, irrespective of shock
velocity, for all but the highest density ($10^{7}$~cm$^{-3}$) gas. We
observe $T_{\rm rot}\approx T_{\rm vib}$ within errors, so that our
data are consistent with the high density C-shock models. Taking
$n=10^{7}$~cm$^{-3}$, then our line ratios constrain the shock
velocity to $\approx 25$~km~s$^{-1}$ for a planar shock and $\approx
30$~km~s$^{-1}$ for a bow shock. These velocities result in gas
temperatures of $\approx 2000$~K, again as observed. This argument is
somewhat contingent on our assumption that all of the hydrogen is in
molecular form. If we allow 20 per cent of the hydrogen nuclei to be
in atomic form, then similar rotational and vibrational temperatures
are achieved for all densities ($10^{4}-10^{7}$~cm$^{-3}$) in our
models. However, as mentioned above, there is evidence for densities
$\sim 10^{7}$~cm$^{-3}$ in the cavity walls, where material is swept
up. Under these conditions hydrogen molecules should (re)form on dust
grains on a timescale of $<100$~yr (Smith et al. 2003), so the assumption 
of fully molecular hydrogen in the dense pre-shock gas seems justified.

To summarize, our line ratios indicate that the H$_2$ emission arises
in C-shocks in dense ($>10^{7}$~cm$^{-3}$) gas. At these densities and
with a transverse magnetic field of 4~mG, a shock velocity in the
range $25-30$~km~s$^{-1}$ is consistent with observation.

\section{Distribution of H$_2$ emission}
\label{h2dist}
\begin{figure}
\epsfxsize=8.5cm \epsfbox[10 -35 262 628]{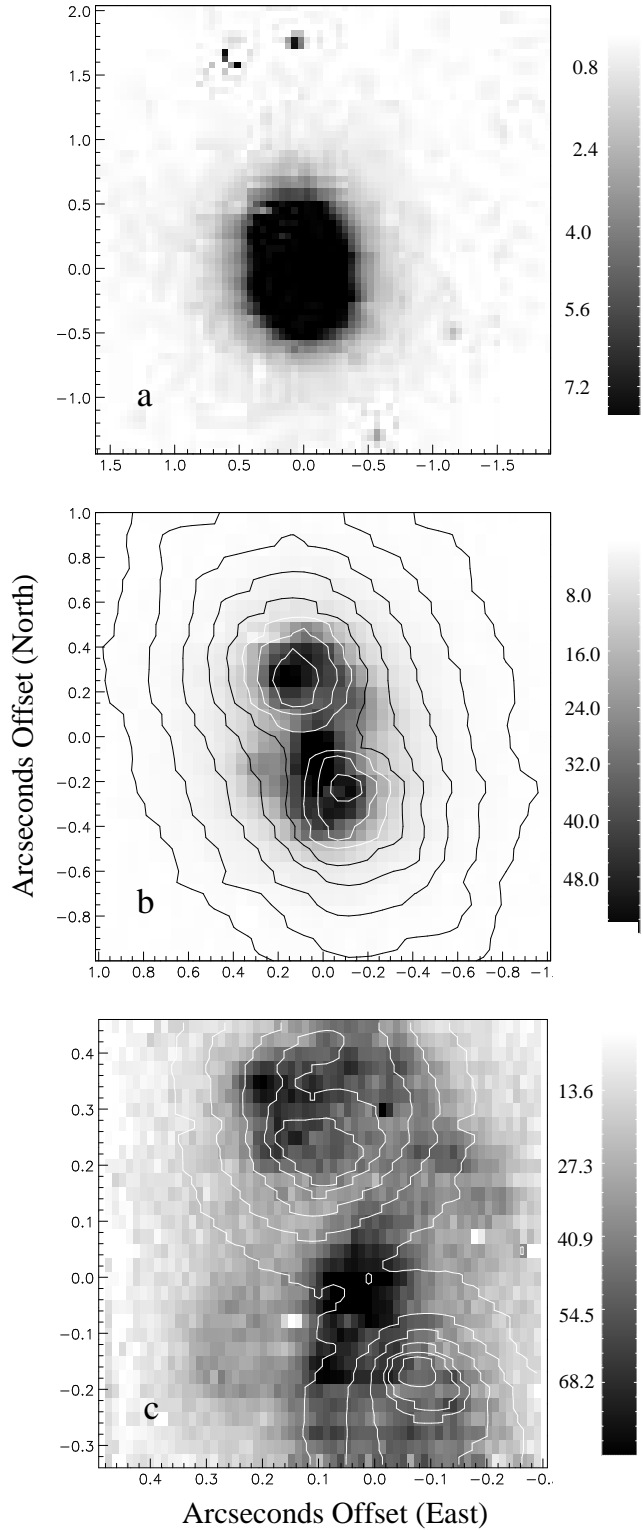}
\vspace{-2cm}
\caption{Images of the $1-0$ S(1) emission in a format similar to that
of the {\it K}-band continuum emission (Fig.~1). The entire field, showing
the faint emission, is seen in the MRM image in {\it (a)} with the
brighter central emission shown on an expanded spatial scale in {\it
(b)}. The HRM $1-0$~S(1) image is shown in {\it (c)}. Greyscale
units are $\times10^{-18}$~W~m$^{-2}$~arcsec$^{-2}$. The H$_2$ emission
in {\it (b)} and {\it (c)} is superimposed with contours of the MRM
and HRM {\it K}-band continuum emission, respectively, with contour levels
identical to those in Fig.~1.}
\end{figure}

The spatial distribution of the H$_2$ emission lines can be
investigated by collapsing the continuum-subtracted datacube over the
width of a line, to form an image of the line emission. This is shown
for the brightest H$_2$ line, the $1-0$~S(1) line, in
Fig.~6.  The format is the same as for Fig.~1, which shows
the distribution of the continuum emission.  Firstly, in
Fig.~6a, we see that the field star to the north-west of
IRAS 18276 has disappeared, as expected if the continuum subtraction
is successful. Secondly, the H$_2$ emission appears more centrally
concentrated than the {\it K}-band continuum (Fig.~1), and
not as elongated. There is no evidence for the searchlight beams, seen
clearly in the continuum, confirming that these are scattered light 
features.
At higher brightness contrast (Fig.~6b) we clearly see
H$_2$ emission emanating from the vicinity of the bipolar lobes, but
we also see emission in the equatorial region between the lobes, where
the continuum contours pinch in due to the high optical depth of the
circumstellar torus. In particular, there is a bright spot of H$_2$
emission located close to the centre of this region.  The distribution
of H$_2$ is therefore quite different from that of the scattered
light. The other $1-0$~S- and Q-branch lines show a very similar
brightness distribution, with a central bright spot, emission from the
lobe regions and fainter emission from the equatorial region. The
$2-1$~S-branch lines (not shown) have the same overall appearance as
the $1-0$ lines. We have analysed the $1-0$~S(1)/$2-1$~S(1) flux ratio
in the lobe regions and in the equatorial regions and find no
significant difference within errors, suggesting that the shock
characteristics are similar in these regions.

The HRM image of the $1-0$ S(1) emission over a field of view of
$0.8\times0.8$ arcsec is shown in Fig.~6c, with HRM continuum contours
superimposed. At this spatial resolution (FWHM $\sim 70$~mas) the
H$_2$ emission in the nebular lobes is seen to be clumpy, with a patch
of emission to the north-east of the continuum peak in the north
lobe. In the south lobe, the brightest H$_2$ emission runs down the
east side of the lobe, again avoiding the continuum peak. This region
coincides with the reddest colours in the $M_s-L_p$ map of SC07 (their
fig.~11) which suggests that the H$_2$ emission arises within the
dense shocked interface layer between the cavity and the surrounding
envelope. This interpretation is consistent with the high-density
C-shock models presented in Section~\ref{shocks}.

The differentiation between the scattered light and H$_2$ emission is
most noticeable in the equatorial regions. We see a bright, diffuse
patch of H$_2$ emission in the centre of the obscuring lane
(Fig.~6c). The circumstellar optical depth to the star
along this line of sight at 2.12~$\umu$m is estimated to be
$\tau_{2.12}>12.3$ (SC07), so that the H$_2$ emitting region must be
in front of the star. Its appearance is again similar to the shape of
the $M_s-L_p$ colour contours, suggesting that the emission arises
from shocks within the walls at the base of the cavity lobes, which
are only just starting to be penetrated by stellar photons in the
$M_s$ image (SC07).

Surrounding the central patch, we see an elliptical or ring-like
region of H$_2$ emission, with a semi-major axis of 0.29 arcsec,
oriented at PA 112\degr. The orientation is similar to that of the
faint equatorial structure seen in the $L_p$-band image of SC07,
radius 0.6~arcsec, which they interpret as the outer parts of an
equatorial torus. These authors also fit the SED with a two-shell
dust model which has a density discontinuity at a radius of $1.6 \times
10^{16}$~cm from the star, corresponding to a sudden drop in mass-loss
rate from $\sim 2 \times 10^{-3}$ to $\sim 3 \times
10^{-5}$~M$_{\odot}$~yr$^{-1}$ at the end of the AGB. The distance to
IRAS 18276 is uncertain, but OH maser phase-lag measurements combined
with VLA interferometry (Bowers, Johnston \& Spencer 1983) indicate
that it lies between 3.4 and 5.4~kpc. Assuming a distance of 4~kpc to
IRAS 18276 (Section~\ref{bipolar}), then this density discontinuity
has an angular offset of $\approx 0.27$~arcsec from the star, which
coincides with our equatorial H$_2$ emission. We propose that this
ring of emission is produced in a shock at the inner boundary of the
AGB envelope.

If we assume that the deprojected equatorial H$_2$ distribution is in fact
circular, then the measured eccentricity of 0.86 corresponds to a tilt
of the polar axis of this equatorial structure by $\approx
22$\degr. relative to the plane of the sky. The PA of this axis, at
25\degr, lies close to the bipolar axis defined by the continuum peaks
(PA = 24\degr; Section~3.1) and to the illumination axis determined
from near-infrared polarimetry (PA = 23\degr; Gledhill 2005).

\section{CO bandheads} 
\label{CO}

\begin{figure*}
\epsfxsize=15.5cm \epsfbox[-190 -380 363 0]{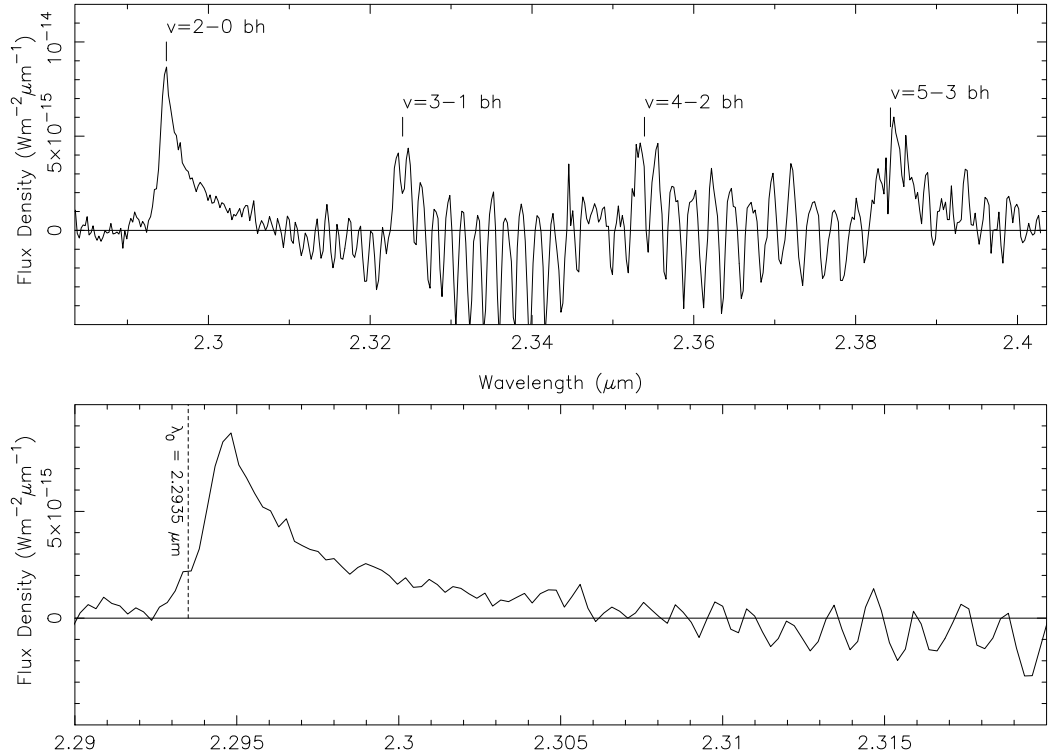}
\caption{The continuum-subtracted spectrum in the first CO overtone
region showing the first 4 bandheads in emission (upper) with an expanded 
view of the ${\rm v}=2-0$ bandhead (lower).}
\end{figure*}

Between 2.29~$\umu$m and the H$_2$ Q-branch transitions longward of
2.4~$\umu$m, the {\it K}-band spectrum is dominated by the first
overtone ro-vibrational transitions of CO. In Fig.~7 we
show the continuum-subtracted spectrum in this region\footnote{Note
  that the CO spectrum most likely extends into the region of H$_2$
  Q-branch emission (the CO ${\rm v}=6-4$ bandhead lies very close to
  the $1-0$~Q(2) line for example), so that the continuum subtraction
  can only be approximate.}.  The ${\rm v}=2-0$ bandhead is clearly in
emission, with evidence for emission in the ${\rm v}=3-1$, $4-2$ and
$5-3$ bandheads also. Beyond $\sim 2.31~\umu$m the individual ${\rm
  v}=2-0$ rotational transitions are resolved in our data and are seen
to go into absorption (roughly at J = 23). We estimate the emitted
flux in the $2-0$ bandhead between 2.293 and 2.306~$\umu$m to be
$4.66\pm0.37\times10^{-17}$~W~m$^{-2}$. The peak of the bandhead lies
at 2.2948~$\umu$m, which is a shift of $1.315\times10^{-3}~\umu$m
redward of the rest wavelength, $\lambda_o=2.2935~\umu$m. The lower J
(${\rm v}=2-0$ R-branch) transitions (in absorption) extend across the
${\rm v}=3-1$ bandhead [R(0) is at $\sim 2.347~\umu$m in our data] so
that it is not possible to reliably estimate the flux in this
bandhead. Similarly, the P-branch extends across the ${\rm v}=4-2$ bandhead
\footnote{The CO ${\rm v}=4-2$ bandhead also includes the H$_2$
  $2-1$~S(0) line at $\lambda_o=2.3556~\umu$m. However, assuming a gas
  temperature of $T_{\rm g}=2053$~K (Table 3), this would be a weak
  feature with an estimated flux of $5.64\times10^{-19}$~W~m$^{-2}$.}.

The continuum-subtracted datacube has been collapsed in wavelength
between 2.2926 and 2.3061~$\umu$m to
form an image of the emission in the ${\rm v}=2-0$ bandhead, which is
shown in Fig.~8 superimposed with contours of the {\it
  K}-band white light image from Fig.~1b. Unlike the
distribution of H$_2$ emission, the distribution of CO emission is
indistinguishable from that of the continuum, which is dominated by
scattered light. As the continuum subtraction in this spectral region
is difficult, we have examined the individual wavelength channels in
the continuum-subtracted datacube to ensure that the subtraction is
successful and that the resulting CO image is not contaminated with
continuum emission. We are confident that Fig.~8 shows
the distribution of the CO bandhead emission. The fact that the CO
emission has the same spatial distribution as the scattered continuum
suggests that it too originates close to the central object and is
scattered into our line of sight by dust in the walls of the polar
cavities. We can therefore predict that spectropolarimetry of the CO
feature would reveal it to have the same high degree of linear
polarization as the continuum at that wavelength. We find no
difference in the bandhead profiles between the northern and southern
lobes and in both cases the peak of the ${\rm v}=2-0$ bandhead lies at
$2.2948~\umu$m, red-shifted relative to the rest wavelength.

\begin{figure}
\epsfxsize=9cm \epsfbox[0 -50 485 459]{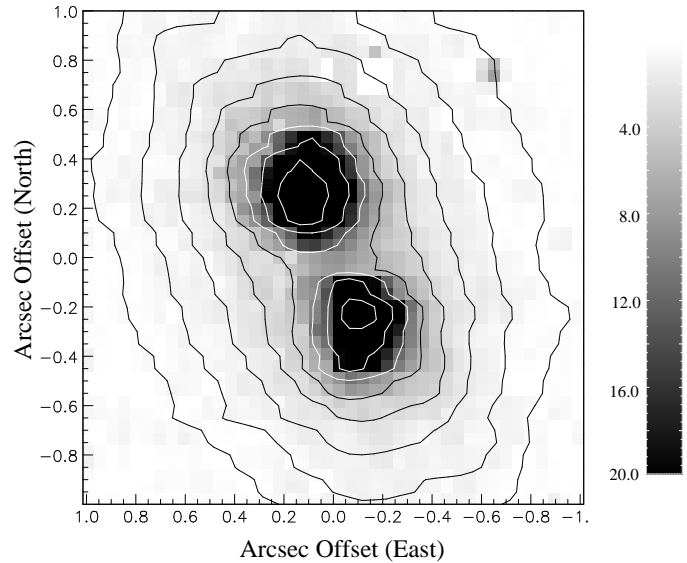}
\vspace{-2cm}
\caption{A continuum-subtracted image of the ${\rm v}=2-0$ bandhead emission
  superimposed with {\it K}-band continuum contours (Fig.~1b), showing
  that the distribution of CO emission is indistinguishable from that
  of the scattered continuum. Greyscale levels are in units of
  $10^{-18}$~W~m$^{-2}$~arcsec$^{-2}$.}
\end{figure}

First overtone CO emission has been widely reported in young stellar
objects (YSOs), where the emission is thought to arise from dense
($n>10^{10}$~cm$^{-3}$) regions of molecular gas at temperatures in
the range 2000-5000~K (e.g. Scoville et al. 1979; Scoville et
al. 1983).  In these sources the gas is likely to reside in a 
circumstellar accretion
disc close to the YSO and modelling of the bandhead profiles and
strengths is in some cases consistent with emission produced in a
Keplerian disc located within a few AU of the star (e.g. Carr 1989;
Bik \& Thi 2004). 

Our results are consistent with the CO overtone emission being
excited in a compact region close to the star. The distribution of the
emission and of the continuum is indistinguishable
(Fig.~8), indicating that both originate from an
unresolved central source. This source is completely obscured from
direct view in the near-IR by the optically thick dust torus and is
instead seen reflected by dust grains in the polar lobes.  The ${\rm
  v}=2-0$ bandhead profile shown in Fig.~7 is remarkably
similar to the profiles observed in a number of YSOs, especially those
modelled with low-inclination (i.e. close to pole-on) circumstellar
discs (e.g. Chandler, Carlstrom \& Scoville 1995; Bik \& Thi 2004). In
these YSO models the CO is collisionally excited in a disc
around the star with temperature in the range 3000-4000~K and
densities $n \ge 10^{10}$~cm$^{-3}$. Radiative excitation mechanisms
(pumping by UV or IR photons) do not yield the observed rotational
distributions or bandhead ratios (Scoville et al. 1979; Scoville,
Krotkov \& Wang 1980; Carr 1989). Comparing our $2-0$ bandhead
spectrum with the model spectra of Scoville et al. 1983, for the BN
object, the best match is for an optically thin line model, with an
excitation temperature of 3500~K and velocity broadening of between 50
and 100~km~s$^{-1}$. However, collisional excitation under these
conditions would be expected to produce a $3-1$ bandhead that is at
least as strong as the $2-0$ bandhead (e.g. Kraus et al. 2000) and
this is generally what is observed in both YSO and post-AGB CO
overtone emitters (see references above). This is clearly not the case
for IRAS 18276, although as mentioned above, the ${\rm v}=3-1$
bandhead is strongly affected by the lower ${\rm v}=2-0$ rotational
transitions which are in absorption, so that the $3-1$ flux appears
much reduced. This effect is also seen to some extent in the BN object
spectrum (Scoville et al. 1983) where cooler gas along the line of
sight is thought to be responsible, and in transient CO emission
observed in the yellow hypergiant $\rho$~Cas (Gorlova et al. 2006).

A further signature seen in YSO spectra, indicating rotating discs, is
the presence of a blue wing or splitting of the bandhead peak into
blue- and red-shifted components, depending on the inclination of the
disc to the line of sight (e.g. Chandler et al. 1995; Najita et
al. 1996). These features are not evident in our data, however, our
view of the central region is provided by dust grains close to the
polar axis so that if a central rotating disc is present (and assuming
that its major axis lies close to the nebular axis) it would be viewed
close to pole-on. We therefore cannot rule out rotation of any central
disc. Similarly, we cannot determine from these data whether the
  gas in the disc is infalling or outflowing, however its presence
  within a few stellar radii of the source is a strong signature of
  ongoing mass loss.

\section{Velocity structure}
\label{velocity}

\begin{figure*}
\epsfxsize=17cm \epsfbox{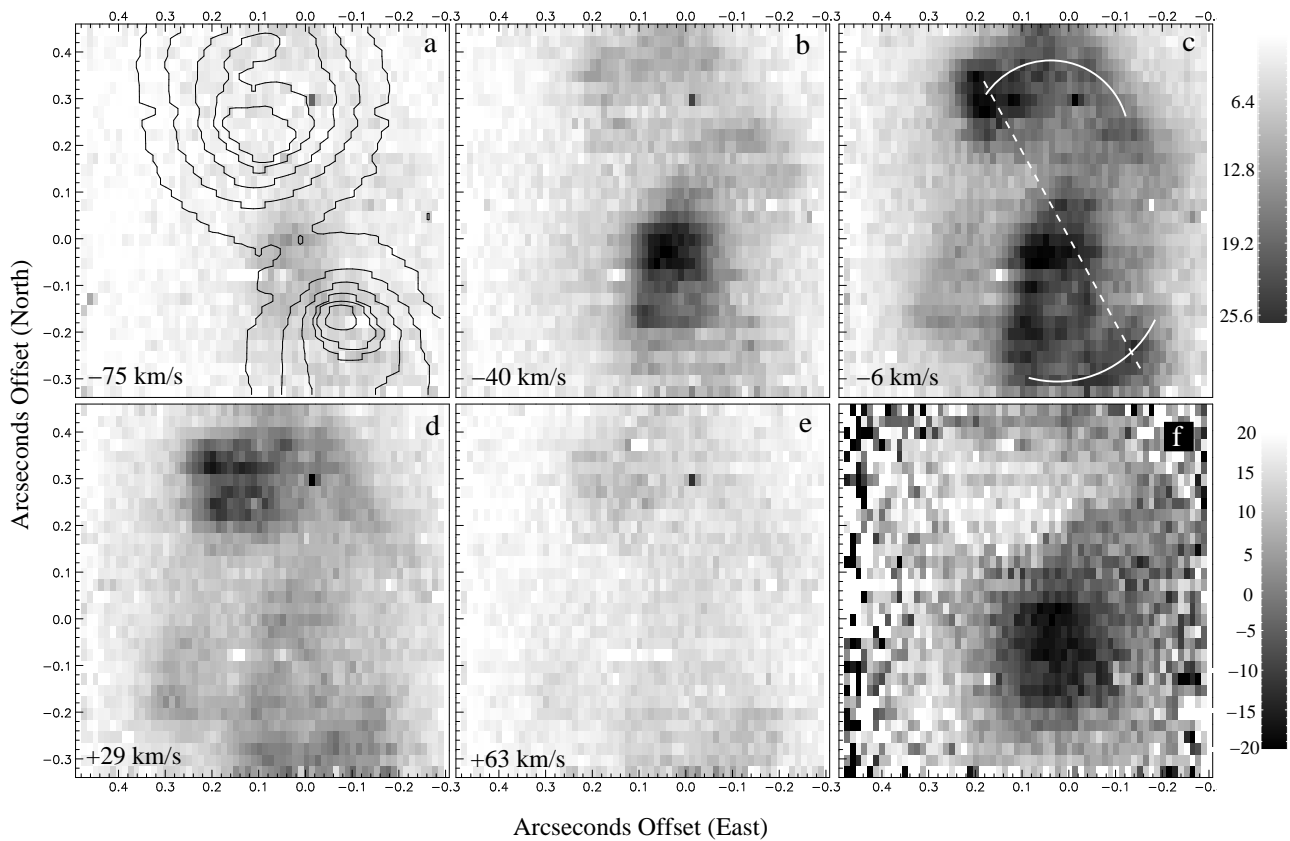}
\caption{Panels {\it (a)} to {\it (e)} show `channel maps' from the
  HRM datacube across the $1-0$ S(1) line, labelled with the velocity
  at the centre of the channel (relative to a systemic velocity of
  61.5~km~s$^{-1}$).  Each image is displayed at the same greyscale
  range as indicated by the scale to the right of panel {\it (c)} in
  units of $10^{-18}$~W~m$^{-2}$~arcsec$^{-2}$. Panel (a) also shows
  {\it K}-band continuum contours. Panel {\it (f)} shows a position-velocity map of the $1-0$ S(1) emission obtained by
  Gaussian-fitting the emission line. The greyscale units are
  km~s$^{-1}$ relative to systemic.  In panel {\it (c)} we also mark
  two arcs of emission along with an axis joining the bright emission
  patch in the northern lobe to an emission patch in the southern
  lobe. These features are discussed further in Section~8.4. }
\end{figure*}

Information on the velocity structure of the molecular gas can be
obtained from the {\sc SINFONI} spectral cubes even though the
velocity resolution is quite coarse, with the width of a spectral
pixel (channel) corresponding to $34.6$~km~s$^{-1}$ at
2.122~$\umu$m. The {\sc SINFONI} user manual quotes a FWHM in the
spectral direction of 2 pixels in the {\it K}-band, although
investigation of the arc lines suggests that the total width of the
spectral profile can be broad, with the line emission spread over as
much as 10 pixels in the wavelength direction.  The accuracy of the
wavelength calibration is a further possible source of error when
determining velocities, so we have checked this by matching
atmospheric absorption features in the science data against a high
spectral resolution atmospheric template to determine and correct for
any offset.

The systemic velocity of IRAS~18276 is $V_{\rm
  LSR}=61.45\pm0.18$~km~s$^{-1}$ as determined from OH maser
observations (Bains et al. 2003) which is in agreement with $V_{\rm
  LSR}=62$~km~s$^{-1}$ determined from the $^{12}$CO $J=1-0$ line
centre (SC07).

\subsection{${\rm v}=2-0$ CO bandhead}
\label{COvelocity}

The ${\rm v}=2-0$~$^{12}$CO bandhead peak lies at 2.9848~$\umu$m
relative to the rest wavelength of 2.2935~$\umu$m, equivalent to a
LSR-corrected velocity of $V_{\rm LSR}=156\pm15$~km~s$^{-1}$ or
$V_{\rm rel}=95\pm15$~km~s$^{-1}$ relative to the systemic velocity of
61.45~km~s$^{-1}$. The error on our velocity corresponds to half a
spectral pixel at the wavelength of the CO bandhead, so is a
conservative estimate. As mentioned in Section~\ref{CO}, the
distribution of CO emission is indistinguishable from that of the
scattered continuum, and we measure the same red-shifted relative
velocity in both the north and south lobes. We interpret this red
shift as the radial velocity (from the star), along the polar axis, of
the dust grains responsible for scattering the CO emission. In other
words, assuming that the bandhead emitting region is an unresolved,
centrally located source, obscured from direct view (as described in
Section~\ref{CO}) moving at the systemic velocity, we can use the peak
emission wavelength of the reflected bandhead to estimate the velocity
of the dust in the axial outflow, $V_{\rm axial}=
95\pm15$~km~s~$^{-1}$.

\subsection{H$_2$ emission}

As noted earlier, the distribution of H$_2$ emission differs
considerably from that of the scattered continuum, and its velocity
structure can be investigated by examining slices through the SINFONI
datacube (channel maps). Fig.~9, panels (a) to (e), shows 5 slices
through the $1-0$ S(1) line taken from the HRM cube, representing
  contiguous 34 km~s$^{-1}$-wide spectral pixels, with the 
  central velocity (relative to systemic) shown in each panel
(negative velocities are blue shifted). As mentioned above, the FWHM
in the spectral direction is at least 2 channels, however there is
clearly velocity structure in the emission. Little emission is seen in
the extreme velocity
channels, (a) and (e), and what there is probably results from the
spectral PSF. The central bright patch of H$_2$ emission is prominent
in panels (b) and (c), but absent in (d), showing that this region has
a blue-shifted component. Conversely, the bright patch in the northern
lobe is prominent in panels (c) and (d), but absent in
(b), indicating that it has a red-shifted component. The faint
equatorial structure to the east of the star also appears more red
than blue shifted, as does the southern lobe, as seen at the
lower edge of the field in panel d.  Panel (f) shows a 
position-velocity map of the
emission, generated by fitting Gaussian profiles in the spectral
direction at each spatial pixel using the {\sevensize VELMAP}
application in the STARLINK software collection. This map therefore
shows the velocity of the peak of the emission line at each position,
and the central patch is clearly blue shifted relative to the systemic
velocity, whereas the northern patch, eastern equatorial region and,
to some extent the southern lobe, are red shifted.  The same tendency
for a blue-shifted central patch and red-shifted lobes is seen in all
other $1-0$ S- and Q-branch lines. The effect is harder to see in the
$2-1$ lines as these are intrinsically fainter.

If the central H$_2$ patch lies in front of the star
(Section~\ref{h2dist}) then this is consistent with it being
predominantly blue shifted; the emission in this region results from
shocks in gas flowing mainly towards us. The far side (red-shifted
component) of this outflow is not seen due to the high obscuration
through the central region. The position-velocity diagram in
  Fig.~9f shows a line-of-sight velocity of up to -20~km~s$^{-1}$ in
  this region.  As the orientation of any outflow in this region is
  unknown, it is not possible to estimate the actual outflow
  velocity.

The red-shifted component in the northern
lobe is harder to explain, especially as the relative continuum
brightness of the northern and southern lobes suggest that the
northern lobe is tilted slightly towards the observer (e.g. Gledhill
2005). It is possible that the red-shifted patch lies on the far side
of the northern lobe and so has a line-of-sight velocity component
away from the observer. It seems unlikely, however, that we are
  seeing emission preferentially from the far side of the lobe, given
  that our analysis in Section 4 indicates that the shocks are
  occurring in high density and hence optically thick regions.
Alternatively, some of the H$_2$ emission may be scattered by dust
grains flowing along the cavity, away from the H$_2$ emitting region,
adding a red-shifted component to the velocity of the line. This
  would also impart a similar red shift to the emission from the
  southern lobe, as observed, so this would seem a viable
  explanation. To investigate these possibilities further will
require observations with higher velocity resolution than that
afforded by {\sc SINFONI}. Also, polarimetry of the H$_2$ lines would
help to determine whether there is a scattered component to the
emission.

\section{Further discussion} 

\subsection{The central regions}
Bedijn (1987) found that in order to fit the infrared SED of IRAS
18276 it was necessary to assume that mass loss did not cease abruptly
at the end of the AGB, but instead continued at a decreasing rate. This
leads to a region of warm dust interior to the AGB shell causing the
excess flux seen in $2-6~\umu$m photometry. Emission from warm dust is
also indicated by our {\it K}-band spectrum (Fig.~2) in which the
continuum flux density rises by a factor of 2 between 1.95 and
2.45~$\umu$m.  The SED fit of SC07 indicates a stellar temperature and
luminosity of $T_{\rm eff}=7000$~K and $L=6400$~L$_{\odot}$ so that
the photospheric emission would be expected to fall over this
wavelength range. Their model also indicates that the continuum is
  composed of 20 per cent stellar light and 80 per cent emission from
  a warm circumstellar dust cloud and that both components suffer an 
extinction of
  $\tau_{2.12~\mu{\rm m}}\sim 3.7$. Taking these assumptions, our
  {\it K}-band continuum slope implies a blackbody temperature of 
    $\sim450$~K for the warm dust, which is within the range of
$300 - 500$~K determined by SC07.


An alternative location for the dust could be in the lobes of the
cavities, where we see bright continuum emission in our images, but it
would not be possible for dust grains to achieve temperatures of 
  $\sim 450$~K in equilibrium with the stellar radiation field at
these offsets. Our H$_2$ imaging shows that shocks are being driven
into the cavity walls with gas temperatures $\sim 2000$~K, so it is
possible that dust could be heated by the shocks. However, the
distribution of shocked H$_2$ emission is very different to that of
the {\it K}-band continuum, which is also highly polarised and
therefore scattered.  The most logical explanation is that the 
  warm dust exists in a compact central region, which may
  extend all the way in to the condensation radius at $\sim
10^{14}$~cm and that mass loss is ongoing.

This conclusion is reinforced by our observation of CO bandhead
emission which reveals the existence of hot ($T>2000$~K) and dense
($n>10^{10}$~cm$^{-3}$) molecular gas in a compact central region. The
most likely location for this collisionally excited gas is close to
the star, and typically within a few stellar radii (e.g. Oudmaijer et
al. 1995; Scoville et al. 1983), in an inner torus or wind. A distance
of 4$R_*$ corresponds to $1.5\times 10^{13}$~cm and is again
consistent with extrapolation of the gas temperatures and densities of
the SC07 model inwards (their fig.~13). We also see absorption in the 
lower J lines of the ${\rm v}=2-0$ CO transition (Fig.~7) which we
interpret as a region of cooler gas, lying between the hotter central
regions and the observer. As the radiation from the central region
is being scattered into our line of sight by dust in the bipolar lobes, 
this cooler gas is probably part of the general mass outflow along 
the bipolar axis.

We also note in Fig.~2 the detection of the Na~{\sevensize I}
  doublet in emission. This is further evidence for the presence of
  high-density gas in the central region. As in the case of the CO
  bandhead emission, the Na~{\sevensize I} line has the same spatial
  distribution as the continuum indicating that it is scattered.  The
  Na~{\sevensize I} lines have been observed in pre-main sequence jet
  sources associated with Herbig-Haro Objects (e.g. HH34 IRS and HH26
  IRS) particularly when CO bandhead emission is also present
  (Antoniucci et al.  2008). These authors suggest that the
  Na~{\sevensize I} emission may originate from low-ionization gas
  that lies closer to the star (and is hotter) than that responsible
  for the CO emission.

A photospheric temperature of $T_{\rm eff}=7000$~K is at odds with our
non-detection of Br$\gamma$. Many F and G spectral type post-AGB stars
show a prominent Br$\gamma$ absorption line (e.g. Hrivnak, Kwok \&
Geballe 1994), the line only disappearing completely for spectral types later
than about mid-K (Wallace \& Hinkle 1997). This may argue for a much
cooler effective temperature for IRAS~18276, but it seems more likely
that any stellar absorption features are obliterated by the strong
{\it K}-band dust continuum.

\subsection{CO bandhead emission from evolved stars}

First overtone bandhead emission from CO has been noted in a number of
evolved stars. {\it K}-band spectroscopy of 16 post-AGB targets by Hrivnak,
Kwok \& Geballe (1994) detected emission in 3 objects
(IRAS~19114+0002, IRAS~22223+4327 and IRAS~22272+5435) which they
suggest is circumstellar in origin. In the case of IRAS~22272+5435,
the CO bandheads changed from emission to absorption within a 3 month
period, which Hrivnak et al. (1994) interpret as evidence for sporadic
mass loss. These three objects are known to be associated with
multi-polar nebulosity revealed in scattered light at optical and
infrared wavelengths (e.g. Ueta et al. 2000, Gledhill et al.  2001)
which again suggests a link with time-varying outflows.

Garc\'{i}a-Hern\'{a}ndez et al. (2002) detect CO emission in three
post-AGB objects: IRAS~08005-2356, IRAS~17423-1755 and
IRAS~10178-5958. IRAS~08005-2356 has limb brightened bipolar lobes in
the {\it V}-band (Ueta et al. 2000). Spectroscopic observations of the
H$\alpha$ and H$\beta$ lines show P-Cygni profiles which are
interpreted as a fast (up to 400 km~s$^{-1}$) wind from the star
(Slijkhuis, Hu \& de Jong 1991). These authors also report a
near-infrared excess which is modelled as emission from hot dust close
to the star. {\it V}- and {\it I}-band images of IRAS~17423-1755 (Ueta
et al. 2000) show a remarkable S-shaped nebula with a pronounced
central obscuring disc. A series of collimated axial jets and emission
line knots (Borkowski, Blondin \& Harrington 1997) provide evidence
for episodic and on-going mass-loss (Riera et
al. 2002). IRAS~10178-5958 is a highly collimated bipolar object with
open lobes, a central obscuring disc and prominent H$_2$ emission
(e.g. Hrivnak et al. 2008).

CO bandhead emission is reported in the post-AGB objects HD~170756
(AC~Her) and HD~101584 (IRAS~11385-5517) by Oudmaijer et al. (1995)
who also suggest that the emission originates either in a
circumstellar disc or a dense outflow close to the star. CO bandheads
are seen in emission in the bipolar pre-PN M~1-92 (Hora,
Latter \& Deutsch 1999; Davis et al. 2003) in which H$_2$ emission is
seen from Herbig-Haro-like bow shocks along the bipolar jet axis
(Bujarrabal et al. 1998). 

It seems clear that CO bandhead emission is closely linked to
on-going mass loss and the presence of a circumstellar disc. In some
cases there is direct evidence for this mass loss in the form of jets
and fast outflows (e.g. IRAS~17423-1755). However, many bipolar
  pre-PN do not show prominent CO features in the near-IR.  AFGL 2688
  (The Egg Nebula) is often considered to be a prototypical example of
  a pre-PN and shares many characteristics with IRAS 18276 (see
  Sec.~8.3) but CO bandheads have not been detected (Davis et
  al. 2003). IRAS 17150-3224 is a well-studied bipolar pre-PN and
  shows emission from shocked H$_2$ in the near-IR, but not from CO
  (Davis et al. 2003). We suggest that CO bandhead emission in pre-PN
  indicates an ongoing intense phase of outflow or jet activity linked
  with the presence of shocks in dense ($\sim 10^{10}$~cm$^{-3}$)
  molecular gas. Objects not exhibiting CO bandheads may be in a more
  quiescent phase.

\subsection{Equatorial H$_2$ emission}

The occurrence of equatorial H$_2$ structures in post-AGB objects is,
so far, uncommon. The most notable example is the case of AFGL 2688,
in which {\em HST} NICMOS observations show an
equatorial region of $1-0$~S(1) emission comparable in extent with the
size of the bipolar lobes at that wavelength (Sahai et al. 1998). The
H$_2$ emission in AFGL 2688 also results from shock excitation with a
line ratio of $1-0$~S(1)/$2-1$~S(1) = 11 averaged over the object
(Hora \& Latter 1994), equivalent within errors to our value of
9.4$\pm$3.2 for IRAS 18276. The shock nature and appearance of the
H$_2$ structures in AFGL 2688 imply that high velocity gas makes its
way out through the central dust cocoon, which blocks stellar photons
in the {\it K}-band, to interact with the slower moving gas in the
circumstellar envelope.  Sahai et al. (1998) suggest that this higher
velocity outflow may occur along channels in the dust cocoon. 

In IRAS~19306+1407, a more evolved object with a B-type central star,
the nebulosity has a bipolar configuration in the optical, with a
denser equatorial torus becoming prominent in near-IR polarized light
(Lowe \& Gledhill 2007). These authors model the SED and scattered
light using an axisymmetric envelope with equator-to-pole density
contrast 7:1. H$_2$ emission is seen in the bipolar lobes of this
object as well as in two arcs on either side of the star, coincident
with the inner edges of the dust torus. The emission is best fitted by
a combination of shock- and UV-excitation and there is evidence from
the clumpy nature of the arcs that the wind is beginning to erode
paths through the circumstellar torus (Lowe \& Gledhill 2006).

The ring of equatorial H$_2$ emission in IRAS~18276 also appears to be
produced in shocks at the inner edge of the denser AGB envelope, where
the density drops by two orders of magnitude. In this object the
density step is not yet visible in scattered light or thermal dust
images, so that IRAS 18276 is perhaps an antecedent of the more
evolved IRAS 19306+1407.

\subsection{The bipolar outflow}
\label{bipolar}

The angular extent of the northern lobe, measured from the star to the
polar `cap' (Fig.~1c) is 0.625 arcsec (12.5 milliarcsec pixels).
Assuming $D=4$~kpc (Section~\ref{h2dist}) and that the bipolar outflow
lies in the plane of the sky, then the linear extent of the northern
lobe is $3.7 \times 10^{14}$~m. An axial wind with velocity
95~km~s$^{-1}$ would take 125~yr to travel this distance, which
provides a lower limit for the age of wind-carved bipolar
cavities. The radial distribution of the dust density in the envelope
around IRAS 18276 (modelled by SC07) indicates a period of increasing
mass loss (from $\sim 2 \times 10^{-4}$ to $\sim 2
\times10^{-3}$~M$_{\odot}$~yr$^{-1}$) commencing more than 2300 and
ending $\sim 300$ years ago. This abrupt end to the heavy mass-loss
phase signals the start of post-AGB evolution.
Although uncertainties of a factor of 2 or more may be present
in our estimate of the cavity lifetime, it seems certain that the
bipolar cavities are young relative to the AGB envelope and that they
started to form very soon after the heavy mass-loss phase terminated.

The patchy distribution of the H$_2$ emission was noted in
Sec.~\ref{h2dist}, and seems brightest close to the walls and tips of
the bipolar cavities, forming arcs of emission at the ends of the
cavities (marked on Fig.~9c). These arcs may trace the shock
fronts where the axial outflow hits the ends of the cavities, which
appear to be still closed in the continuum images (which trace the
dust).  The shock characteristics are indicative of gas densities of
the order $n=10^7$~cm$^{-3}$ which are typical of those in the cavity
walls and tips (Sec.~\ref{shocks}), so that this scenario seems
plausible.  The bright patch in the northern lobe is particularly
noticeable. It is quite possible that the material in the cavity walls
has a clumpy distribution, and that impingement of the axial wind on
these clumps leads to the patchy nature of the emission in IRAS 18276;
H$_2$ emission from clumpy structures is seen in the equatorial region
of IRAS 19306+1407 (Lowe \& Gledhill 2006) and spectacularly in the
Helix planetary nebula (NGC~7293; Matsuura et al. 2009). Additionally,
if the axial outflow is more confined and jet-like, it may result in
H$_2$ `emission spots' at the jet tips. We propose that
  such a collimated axial outflow or jet is operating and is currently
  aligned with the emission spot in the northern lobe. If this jet is
  precessing and has excavated the cavities then the arcs of emission
  may represent the recent locus of the jet tip. In Fig.~9c we mark a
possible jet axis, drawn between the bright H$_2$ patch in the
northern lobe and a fainter patch in the southern lobe.  This axis
lies at PA = 29\degr, which is inclined relative to the major axis of
the cavities at PA = 24\degr, defined by the two {\it K}-band
continuum peaks, and to the {\it K}-band polarimetric axis, PA =
$22\pm2$\degr (Gledhill 2005). The angle subtended at the star by the
H$_2$ patch in the northern lobe is $24\pm5$\degr.

\section{Conclusions}

Integral field spectroscopic observations of
IRAS~18276-1431 are analyzed to determine the distribution,
excitation and kinematics of H$_2$ and CO emission in the {\it
  K}-band.  We summarize our conclusions as follows:

\begin{enumerate}
\item We detect the {\it K}-band ${\rm v}=1-0$ and ${\rm v}=2-1$
  S-branch lines of H$_2$, as well as the ${\rm v}=1-0$ Q-branch. The
  $1-0$~S(1)/$2-1$~S(1) ratio is $9.3\pm3.2$ over most of the object,
  which is typical of shock excited emission.  Further evidence for
  shock excitation is provided by an ortho-para ratio of $3.02\pm0.18$
  and roughly equivalent vibrational and rotational excitation
  temperatures of $\approx 2000$~K.  A marginal detection of the ${\rm
    v}=3-2$~S(3) line is again consistent with shocks and indicates
  that UV-pumped fluorescence does not contribute significantly to the
  H$_2$ emission. This accords with the proposed evolutionary status
  in which IRAS~18276-1431 has recently left the AGB, is now carving
  outflows into the AGB envelope via a fast wind, but is not yet hot
  enough to UV-excite its surroundings.

\item Comparison of the observations with shock models favours
C-shocks, with either a planar or curved (bow) configuration, in
high density ($>10^{7}$~cm$^{-3}$) gas with shock velocities of
$25-30$~km~s$^{-1}$. This is also consistent with the previous
detection of a $\approx 4$~mG magnetic field in the torus
region. 

\item The distribution of H$_2$ emission differs considerably from that
of the scattered continuum and implies a clumpy distribution of
material in the walls and tips of the outflow cavities. We suggest
that the H$_2$ emission in these regions is shock excited by a fast
axial outflow impinging upon dense gas. A
bright patch of emission in the northern lobe may provide evidence for an
inclined axial jet.

\item A bright spot of H$_2$ emission is located between the bipolar
lobes where the scattered continuum is a minimum, and must lie between
us and the star. We interpret this as shocks driven by the wind as it
impacts dense material at the base of the cavities. We also see H$_2$ 
emission in the equatorial region, to the east and west of the stellar 
position, which may form an equatorial ring with radius $\approx 0.3$ 
arcsec. The location corresponds to the inner edge of the AGB envelope,
where the density changes by two orders of magnitude, so that the
emission may arise in shocks in the boundary region between the two
(AGB and post-AGB) winds. 

\item The velocity structure of the H$_2$ lines shows that the
central spot has a blue-shifted component with the peak shifted by
$\approx -20$~km~s$^{-1}$ relative to the systemic velocity. Conversely,
the emission from the bipolar lobes and equatorial region appears
red shifted by a similar amount. We suggest that the blue-shifted
central spot results from approaching shock fronts in material
foreground to the star. The red-shifted lobe emission may be due to a
component of line scattering by dust grains in the axial outflow,
receding from the shock.

\item The {\it K}-band spectrum of IRAS 18276-1431 shows a rising
  continuum and evidence for the presence of warm $\sim 450$~K 
dust, suggesting
  that mass loss and dust formation is ongoing in this object.

\item Further evidence for continuing mass-loss is provided by the first
overtone $^{12}$CO ro-vibrational bandheads, which are observed in
emission, which may signify the presence of a circumstellar disc of
hot CO located close to the star. CO bandhead emission is observed in
several other post-AGB sources, all with evidence of ongoing mass
loss, circumstellar discs and collimated outflows/jets. Detection
of the Na~{\sevensize I} doublet in emission further suggests the
presence of high-density gas close to the star.

\item The CO bandhead emission has the same spatial distribution as the
scattered {\it K}-band stellar continuum, indicating that it is also
scattered by dust grains in the polar lobes. This further supports a
central, unresolved, location for the CO emitting region. We use the
red shift of the peak of the scattered ${\rm v}=2-0$ bandhead to estimate
the radial velocity (from the star) of the dust grains responsible for
the scattering, leading to an estimate for the axial velocity of dust
in the outflow of $95\pm15$~km~s$^{-1}$. This in turn places a lower
limit of 125~yr on the age of the bipolar cavities meaning that any
fast wind must have turned on very soon after the cessation of AGB
mass loss.
\end{enumerate}

The overall picture of IRAS 18276 is one in which the central star has
recently moved into the post-AGB phase. Mass loss continues in the
form of a more tenuous and faster wind, which is now driving shocks 
in the surrounding molecular material. Further monitoring of the $K-$band 
spectrum will be useful to determine
if the CO bandhead emission persists, which may indicate a stable region
of hot molecular gas close to the star. 

\section*{Acknowledgments}
This work is based on observations with ESO telescopes at Paranal
Observatory.  We thank the staff of Paranal Observatory for assistance
with these observations. The anonymous referee is thanked for their 
valuable comments.

\end{document}